\documentclass[12pt,a4paper]{article}%
\usepackage{amsmath}%
\usepackage{amsfonts}%
\usepackage{amssymb}%
\usepackage{graphicx}
\usepackage{latexsym}
\usepackage{amsopn}
\usepackage{latexsym}
\usepackage[latin1]{inputenc}
\providecommand{\U}[1]{\protect\rule{.1in}{.1in}}
\newtheorem{theorem}{Theorem}
\newtheorem{acknowledgement}[theorem]{Acknowledgement}

\newcommand{\parenthnewln}{\right.\\&\left.\quad\quad{}}

\begin{document}
\clearpage

\begin{center}
Hamiltonian Formulation of Mimetic Gravity
\end{center}
\vspace{0.5cm}
\begin{center}
O. Malaeb$^{1}$ \\
Physics Department, American University of Beirut, Lebanon
\end{center}
\vspace{3cm}
%

\begin{abstract}
The Hamiltonian formulation of Mimetic Gravity is formulated. Although there are two more equations than those of general relativity, these are proved to be the constraint equation and the conservation of energy-momentum tensor. The Poisson brackets are then computed and closure is proved. At the end, Wheeler-DeWitt equation was solved for a homogeneous and isotropic universe. This was done first for a vanishing potential where agreement with the dust case was shown, and then for a constant potential.
\end{abstract}

\clearpage

\section{Introduction}

Not long ago, Chamseddine and Mukhanov \cite{mimetic} proposed a theory of mimetic dark matter by modifying Einstein's general relativity. Their theory was shown to be a conformal extension of Einstein's theory of gravity. They defined a physical metric in terms of an auxiliary metric, $\tilde{g}_{\mu \nu}$, and first derivatives of a scalar field $\phi$. Interestingly, the equations of motion differ from Einstein equations by the appearance of an extra mode of the gravitational field which can reproduce Dark Matter. An equivalent formulation of the model was given in \cite{golovnev} where the action is written without the auxiliary metric. Then the ghost free models of this theory were discussed in \cite{barvinsky}. It was found that the theory is free of ghost instabilities for a positive energy density. In a recent paper \cite{cosmo}, the mimetic dark matter theory was extended by adding an arbitrary potential and then cosmological solutions were studied. It was shown how various cosmological solutions can be found by choosing an appropriate potential.\paragraph{}
In this paper we are going to consider the canonical formulation of this theory. There are advantages of setting up the Hamiltonian dynamics for this new model such as quantization. Arnowitt, Deser and Misner constructed a canonical formulation of gravity (ADM) in 1959 (\cite{dynamics}). The tetrad form of this ADM formalism was derived in \cite{DeserIsham}. Schwinger constructed an invariant action operator under both local Lorentz and general coordinate transformations \cite{schwinger}. Ashtekar \cite{ashtekar}, \cite{ashtekar1} introduced new variables which lead to simplifications in the gravitational constraints, and his theory lead to loop quantum gravity. Also, the canonical formulations of matter couplings to gravity were considered (see for example (\cite{schwinger1} - \cite{held})).  \paragraph{}
The aim of this paper is to analyze the hamiltonian equations of motion and to check that the degrees of freedom are those of general relativity. Also, Poisson brackets will be computed.

\section{Canonical Form}

%
%

Consider the action of the extended mimetic dark matter theory \cite{golovnev}, \cite{cosmo}
\begin{align}
S ={}& - \int d^4x ({-g})^{1/2} \left(\frac{1}{2} R + \frac{1}{2} \lambda \left( 1 - g^{\mu \nu} \partial_{\mu} \phi \partial_{\nu} \phi  \right) + \textsl{L}_m \right) 
\end{align}
where $\textsl{L}_m$ is an arbitrary potential $V(\phi)$ added. Variation with respect to the variable $\lambda$ gives the constraint equation on the scalar field
\begin{equation}
g^{\mu \nu} \partial_{\mu} \phi \partial_{\nu} \phi = 1.
\end{equation}\newline
To construct the canonical formalism of this theory, we first start by rewriting the action in a $3+1$ dimensional form. The part of the action containing the scalar field $\phi$ becomes 
\begin{align}
S_{\phi} ={}& -  \int d^4x \ \frac{1}{2} N \sqrt{h} \lambda \left( 1 - g^{00} \partial_{0} \phi \partial_{0} \phi - 2g^{0i} \partial_{0} \phi \partial_{i} \phi  + h^{ij} \partial_{i} \phi \partial_{j} \phi \parenthnewln \nonumber
 - \frac{N^{i} N^{j}}{N^2} \partial_{i} \phi \partial_{j} \phi  \right) + N\sqrt{h} V(\phi)
\end{align}
where $g^{00} = 1/N^2$, $g^{0i} = -N^{i}/N^2$ and $g^{ij} = -h^{ij} + N^{i}N^{j}/N^2$ \cite{dynamics}.  \paragraph{}
The momentum conjugate to $\lambda$ is given by
\begin{equation}
p_{\lambda} = \frac{\partial{L}}{\partial \dot{\lambda}} = 0,
\end{equation} 
while that conjugate to $\phi$ is 
\begin{equation}
p = \frac{\partial{L}}{\partial \dot{\phi}} = N \sqrt{h} \lambda \left( g^{00} \partial_{0} \phi +  g^{0i} \partial_{i} \phi \right).
\label{momentum}
\end{equation}  
Therefore, $p_{\lambda} = 0$ is a primary constraint, and this will imply a secondary constraint by demanding its time constancy 
\begin{equation}
0 = \dot{p}_{\lambda} = \left\lbrace p_{\lambda}, H \right\rbrace  = \frac{\delta H}{\delta \lambda} .
\label{plambda}
\end{equation}
By inverting equation (\ref{momentum}), $\dot{\phi}$ can be substituted by the conjugated momentum $p$, and then the Hamiltonian is given by
\begin{align}
H ={}& \frac{N p^2}{2 \sqrt{h} \lambda} + \frac{1}{2} N \sqrt{h} \lambda \left[ 1 + h^{ij} \partial_{i} \phi \partial_{j} \phi  \right] + pN^{i} \partial_{i} \phi + N\sqrt{h} V(\phi).
\end{align}
This Hamiltonian is still a function of the Lagrange multiplier $\lambda$. To exclude it, we solve equation (\ref{plambda}). This gives
\begin{equation}
\lambda = \frac{p}{\sqrt{h} \sqrt{1 + h^{ij} \partial_{i}\phi \partial_{j}\phi}}.
\end{equation}\newline
Upon plugging back we get the final form of the total action
\begin{align}
S = S_g + S_{\phi} {}& = \int d^4 x \  \left( \textsl{L}_{ADM} + p \dot{\phi} - N p \sqrt{1+ h^{ij} \partial_{i} \phi \partial_{j} \phi } - N^{i} p \partial_{i} \phi \nonumber \parenthnewln
- N \sqrt{h} V(\phi) \right) 
\end{align}
where \cite{dynamics}
\begin{align}
\textsl{L}_{ADM} = \dot{h^{ij}}\pi_{ij} - NR^{0} - N^{i} R_{i}.
\end{align}
$R^0$ and $R_i$ are the intrinsic curvatures given by
\begin{align}
{}& R^{0} \equiv - \sqrt{h} \left[ {}^{3} R + h^{-1} \left( \frac{1}{2} \pi^2  - \pi^{ij}\pi_{ij} \right)  \right] \nonumber \\
& R_i \equiv -2 h_{ik} \pi^{kj}_{\ \mid j}
\end{align}
 
\section{Equations of Motion}
  
Starting from the total action constructed above
\begin{equation}
S = \int \dot{h}^{ij} {\pi}_{ij} + p \dot{\phi} - N \left( R^{0} + p\sqrt{h^{ij} \partial_{i} \phi \partial_{j} \phi + 1}  \right) - N^{i} \left(R_{i} + p \partial_{i} \phi  \right) - N \sqrt{h} V(\phi) ,
\end{equation}
the equations of motion are found by varying with respect to the variables. We first obtain
\begin{align}
\dot{h}^{ij} = \left\lbrace h^{ij}, H  \right\rbrace = 2 N h^{-1/2} \left( \pi^{ij} - \frac{1}{2} h^{ij} \pi \right) + N^{i\mid j} + N^{j\mid i}
\label{hijeq}
\end{align}
\begin{align}
\dot{\pi}_{ij} &= \left\lbrace \pi_{ij}, H \right\rbrace = - N \sqrt{h} \left( {}^{3}R_{ij} - \frac{1}{2} h_{ij} \, {}^{3}R \right) + \frac{1}{2} N h^{-1/2} h_{ij} \left(\pi^{mn} \pi_{mn} - \frac{1}{2} \pi^{2} \right) \nonumber \\
& - 2 N h^{-1/2} \left(\pi_{im} \pi^{m}_{j} - \frac{1}{2} \pi \pi_{ij} \right) + \sqrt{h} \left( N_{\mid ij} - h_{ij} N^{\mid m}\ _{\mid m}\right) + \left( \pi_{ij} N^{m} \right)_{\mid m} \nonumber \\
& - N_{i}^{\mid m} \pi_{mj} - N_{j}^{\mid m} \pi_{mi} + \frac{N p \partial_{i} \phi \partial_{j} \phi}{2\sqrt{h^{kl} \partial_{k} \phi \partial_{l} \phi + 1}} - \frac{1}{2} N\sqrt{h}V(\phi) h_{ij}
\label{pieq} 
\end{align}
These equations result from varying $\pi_{ij}$ and $h^{ij}$ respectively. The former, equation (\ref{hijeq}), is independent of the scalar field $\phi$ since the action $S_{\phi}$ is independent of $\pi_{ij}$. However, equation (\ref{pieq}) contains two terms as a function of $\phi$ which are not there for Einstein's gravity.\paragraph{}
\noindent Variation of $N$ and $N^i$ yields four constraint equations given by
\begin{align}
{}& R^{0} + p\sqrt{h^{ij} \partial_{i} \phi \partial_{j} \phi + 1} + \sqrt{h} V(\phi) = \textsl{H}_{grav} + \textsl{H}_{\phi} = 0 \nonumber \\
& R_{i} + p \partial_{i} \phi = \textsl{H}_{i \, grav} + \textsl{H}_{i \, \phi} = 0
\end{align}
There are two more equations of motion for the phase variables $(\phi, p)$. That for the scalar field is given by
\begin{equation}
\dot{\phi} - N \sqrt{h^{ij} \partial_{i} \phi \partial_{j} \phi + 1} - N^{i} \partial_{i} \phi = 0,
\label{pvar}
\end{equation}
while for the conjugated momentum it reads
\begin{equation}
\dot{p} - \partial_{k} \left( \frac{N p \ h^{kl} \partial_{l} \phi}{\sqrt{h^{ij} \partial_{i} \phi \partial_{j} \phi + 1}} + N^{k} p\right) + N\sqrt{h} \frac{d V(\phi)}{d \phi}  = 0
\label{eqofp}
\end{equation}
Therefore, the number of equations of motion are those of Einstein's gravity plus two more equations coming from the variation of the $\phi$ and $p$. However, equation (\ref{pvar}) is exactly the constraint equation, $g^{\mu \nu} \partial_{\mu} \phi \partial_{\nu} \phi = 1$; therefore, it gives no new info. While equation (\ref{eqofp}) is the Bianchi identity as explained below. Therefore, we end up with the same number of equations of motion as those of general relativity. \paragraph{}
Considering only the part of the action depending on the scalar field $\phi$, it is given by
\begin{align}
S_{\phi} ={}& \int d^4 x \  \left( p \dot{\phi} - Np \sqrt{h^{ij}\partial_{i}\phi \partial_{j} \phi + 1} - N^{i} p \partial_{i} \phi  - N\sqrt{h} V(\phi) \right).
\end{align}
The canonical components of stress energy tensor are given by \cite{bojowald}
\begin{align}
{}& T_{00} = - \frac{N}{\sqrt{h}} \left( N \frac{\delta {S_{\phi}}}{\delta N} + 2 N^{i} \frac{\delta {S_{\phi}}}{\delta N^{i}} + 2 \frac{N^{i} N^{j} }{N^2} \frac{\delta {S_{\phi}}}{\delta h^{ij}}\right) \nonumber \\
& T_{0i} = - \frac{N}{\sqrt{h}} \left( \frac{\delta {S_{\phi}}}{\delta N^{i}} + 2 \frac{ N^{j} }{N^2} \frac{\delta {S_{\phi}}}{\delta h^{ij}}\right) \nonumber \\
& T_{ij} = - \frac{2}{N\sqrt{h}} \left( \frac{\delta {S_{\phi}}}{\delta h^{ij}}\right).
\end{align}
Computing the variations, we get
\begin{align}
{}& T_{00} = \frac{N^2 p}{\sqrt{h}} \sqrt{h^{ij} \partial_{i} \phi \partial_{j} \phi + 1} + \frac{2N}{\sqrt{h}} N^{i}p \partial_{i}\phi + \frac{N^i N^j}{\sqrt{h}} \frac{p\partial_i \phi \partial_j \phi}{\sqrt{h^{ij} \partial_{i} \phi \partial_{j} \phi + 1}} \nonumber \\
& \quad \quad + N^2 V(\phi) -N^i N_i V(\phi) \nonumber \\ 
& T_{0i} =  \frac{N}{\sqrt{h}} p \partial_i \phi + \frac{N^j}{\sqrt{h}} \frac{p\partial_i \phi \partial_j \phi}{\sqrt{h^{ij} \partial_{i} \phi \partial_{j} \phi + 1}} - N_i V(\phi) \nonumber \\
& T_{ij} = \frac{p\partial_i \phi \partial_j \phi}{\sqrt{h}\sqrt{h^{ij} \partial_{i} \phi \partial_{j} \phi + 1}} - h_{ij} V(\phi)
\end{align}
This is in agreement with what is found in the Lagrangian formalism, $T_{\mu \nu} = \lambda \partial_{\mu} \phi \partial_{\nu} \phi$, upon using equation (\ref{pvar}). Then starting from
\begin{equation}
\nabla_{\mu} T^{\mu}_{\, \nu} = \frac{1}{\sqrt{-g}} \partial_{\mu} \left( \sqrt{-g} T^{\mu}_{\, \nu}\right)  - \Gamma^{\mu}_{\nu \rho} T^{\rho}_{\, \mu},
\end{equation}
it is easy to show that equation (\ref{eqofp}) is just the identity $\nabla_{\mu} T^{\mu}_{i} = \nabla_{0} T^{0}_{i} + \nabla_{j} T^{j}_{i} = 0$. \paragraph{}
To summarize, the equations of motion for mimetic gravity are those of Einstein's gravity plus two more equations. These are for the scalar field, which is reinterpreted as the conservation of the energy-momentum tensor, and the other is the constraint equation. However, the equations resulting from varying with respect to $h^{ij}$, $N$ and $N^{i}$ are those of pure Einstein's gravity \cite{dynamics} but including extra terms as a function of the scalar field $\phi$. This is how the mimetic dark matter enters the picture.

\section{Poisson Brackets}

In the presence of the scalar field, the combined constraints are 
\begin{equation}
\textsl{H} = \textsl{H}_{grav} + \textsl{H}_{\phi}; \quad \textsl{H}_i = \textsl{H}_{i \, grav} + \textsl{H}_{i \, \phi}.
\end{equation}
where the first is the Hamiltonian constraint and the latter is called the diffeomorphism constraint. This coupling is $\lq$non-derivative$\rq$ since the $H_{\phi}$ does not depend on the gravitational momentum $\pi_{ij}$ \cite{held}.\newline 
For computing the poisson brackets, we consider smeared functions to have well-defined algebraic relationships. Then derivatives by fields are free of delta-distributions. Writing the constraints in smeared form \cite{bojowald} we have \begin{align} 
{}& \textbf{H}[N] = \int d^3 x N(x)\textsl{H}(x) \nonumber \\
& {\textbf{D}}[N^{i}] = \int d^3 x N^{i}(x)\textsl{H}_{i}(x).
\end{align}
To find the Poisson bracket of two Hamiltonian constraints, we start by expanding  
\begin{align}
\left\lbrace \textbf{H} [N_1] , \textbf{H}[N_2]\right\rbrace {}&= \left\lbrace \textbf{H}_{grav} [N_1] + \textbf{H}_{\phi} [N_1] , \textbf{H}_{grav} [N_2] + \textbf{H}_{\phi} [N_2] \right\rbrace \nonumber \\
& = \left\lbrace \textbf{H}_{grav} [N_1] , \textbf{H}_{grav} [N_2]  \right\rbrace + \left\lbrace \textbf{H}_{grav} [N_1], \textbf{H}_{\phi} [N_2] \right\rbrace \nonumber \\
& +\left\lbrace \textbf{H}_{\phi} [N_1] , \textbf{H}_{grav} [N_2]  \right\rbrace +\left\lbrace  \textbf{H}_{\phi} [N_1] , \textbf{H}_{\phi} [N_2] \right\rbrace 
\end{align} \newline
The gravitational Hamiltonian constraint, $H_{grav}$, is free of spatial derivatives of the momentum; i.e. the two Poisson brackets $\left\lbrace \textbf{H}_{grav} [N_1], \textbf{H}_{\phi} [N_2] \right\rbrace$ and $\left\lbrace \textbf{H}_{\phi} [N_1], \textbf{H}_{grav} [N_2] \right\rbrace$ cancel out. Therefore, we get
\begin{align}
\left\lbrace \textbf{H} [N_1] , \textbf{H}[N_2]\right\rbrace {}&= \left\lbrace \textbf{H}_{grav} [N_1] , \textbf{H}_{grav} [N_2]  \right\rbrace + \left\lbrace  \textbf{H}_{\phi} [N_1] , \textbf{H}_{\phi} [N_2] \right\rbrace .
\end{align} \newline
Since the Poisson bracket is given by
\begin{align}
\left\lbrace {\textbf{H}}_{\phi} [{N}_1] , {\textbf{H}}_{\phi} [{N}_2]\right\rbrace {}&= \int d^{3}x \left( \frac{\delta {\textbf{H}}_{\phi} [{N}_1]}{\delta \varphi(x)} \frac{\delta {\textbf{H}}_{\phi} [{N}_2]}{\delta p(x)} - \frac{\delta {\textbf{H}}_{\phi} [{N}_1]}{\delta p(x)} \frac{\delta {\textbf{H}}_{\phi} [{N}_2]}{\delta \varphi(x)} \right) ,
\end{align}
the functional derivatives should be first computed. These are given by 
\begin{align}
{}& \frac{\delta {\textbf{H}}_{\phi} [{N}]}{\delta \varphi(x)} = -\partial_{i}\left( \frac{pN h^{ij} \partial_j \phi}{\sqrt{h^{kl} \partial_k \phi \partial_l \phi + 1}} \right) + N \sqrt{h} \frac{dV}{d\phi} \nonumber \\ 
& \quad \frac{\delta {\textbf{H}}_{\phi} [{N}]}{\delta p(x)} = N \sqrt{h^{kl}\partial_{k} \phi \partial_{l} \phi + 1}
\end{align}
which give upon plugging in the integral
\begin{equation}
\left\lbrace {\textbf{H}}_{\phi} [{N}_1] , {\textbf{H}}_{\phi} [{N}_2]\right\rbrace = \textbf{D}_{\phi} \left[ h^{ij} \left( N_2 \partial_j N_1  - N_1 \partial_j N_2 \right)  \right] 
\end{equation}
Similarly we have 
\begin{align}
{}& \frac{\delta {\textbf{D}}_{\phi}}{\delta \phi} = - \partial_i (pN^i) \nonumber \\
& \frac{\delta {\textbf{D}}_{\phi}}{\delta p} = N^i \partial_i \phi
\end{align}
which give upon integration
\begin{align}
{}& \left\lbrace {\textbf{D}}_{\phi} [{N^{i}}_1] , {\textbf{D}}_{\phi} [{N^{i}}_2]\right\rbrace = \textbf{D}_{\phi} \left[ N^j_1 \partial_j N^i_2 - N^j_2 \partial_j N^i_1 \right].
\end{align}\newline
In computing  $\left\lbrace {\textbf{D}} [N^{i}] , {\textbf{H}} [N]\right\rbrace$, the Poisson bracket  $\left\lbrace {\textbf{D}}_{grav} [N^{i}] , {\textbf{H}}_{\phi} [N]\right\rbrace$ survives since the scalar hamiltonian constraint $H_{\phi}$ depends on the metric $h^{ij}$. Therefore, we have
\begin{align}
\left\lbrace {\textbf{D}} [N^{i}] , {\textbf{H}} [N]\right\rbrace {}&= \left\lbrace {\textbf{D}}_{grav} [N^{i}] , {\textbf{H}}_{grav} [N]\right\rbrace + \left\lbrace {\textbf{D}}_{\phi} [N^{i}] , {\textbf{H}}_{\phi} [N] \right\rbrace \nonumber \\
& + \left\lbrace {\textbf{D}}_{grav} [N^{i}] , {\textbf{H}}_{\phi} [N] \right\rbrace . 
\end{align}
Computing the Poisson bracket for the scalar constraints we get
\begin{align}
\left\lbrace {\textbf{D}}_{\phi} [N^{i}] , {\textbf{H}}_{\phi} [N] \right\rbrace {}&= pN^k \partial_k N \sqrt{h^{ij} \partial_i \phi \partial_j \phi + 1} - N \sqrt{h} \ N^k \partial_k \phi \ \frac{d V}{d \phi} \nonumber \\
& -\frac{N^k_{\ \vert i} \ pN h^{ij} \partial_j \phi \partial_k \phi}{\sqrt{h^{rs} \partial_r \phi \partial_s \phi + 1}}
\end{align}
Adding $\left\lbrace {\textbf{D}}_{grav} [N^{i}] , {\textbf{H}}_{\phi} [N] \right\rbrace$ to it cancels the last two terms and gives the expected result
\begin{equation}
\left\lbrace {\textbf{D}} [{N^{i}}] , {\textbf{H}} [N]\right\rbrace = \textbf{H} \left[ N^{i} \partial_{i} N \right]. 
\end{equation}
Therefore the full constraint algebra is given by 
\begin{align}
{}& \left\lbrace {\textbf{H}} [{N}_1] , {\textbf{H}} [{N}_2]\right\rbrace = \textbf{D} \left[ h^{ij} \left( N_2 \partial_j N_1  - N_1 \partial_j N_2 \right)  \right] \nonumber \\
& \left\lbrace {\textbf{D}} [N^{i}_1] , {\textbf{D}} [N^{i}_2]\right\rbrace = \textbf{D} \left[ N^j_1 \partial_j N^i_2 - N^j_2 \partial_j N^i_1 \right] \nonumber \\
& \left\lbrace {\textbf{D}} [{N^{i}}] , {\textbf{H}} [N]\right\rbrace = \textbf{H} \left[ N^{i} \partial_{i} N \right] 
\end{align}
which is equivalent to the Dirac algebra \cite{dirac2} (see also \cite{witt}). 



\section{Wheeler-DeWitt}

To discuss quantum cosmology, the ADM Hamiltonian formulation of general relativity is employed. In this section, we apply the above Hamiltonian method to the particularly simple case of a homogeneous and isotropic universe. The line element is then given by Friedmann-Lematre-Robertson-Walker (FLRW)
\begin{equation}
ds^2 = N(t)^2 dt^2 - a(t)^2 \left( \frac{dr^2}{1 - kr^2} + r^2 \left( d\theta^2 + sin^2\theta d\phi^2 \right)  \right).
\end{equation}
For this isotropic metric, the gravitational action is given by 
\begin{equation}
{\mathcal{S}}_{grav} = 3 \int dt \left( \frac{a \dot{a}^2}{N} - kaN \right).
\end{equation}
Momenta are derived in the usual way as
\begin{equation}
p_a = \frac{\partial {\mathcal{L}}_{grav}}{\partial \dot{a}} = 6 \frac{a \dot{a}}{N}, \quad p_N = \frac{\partial {\mathcal{L}}_{grav}}{\partial \dot{N}} = 0. 
\end{equation}
and the gravitational action is then given by
\begin{align}
\mathit{S}_{grav} {}& = \int d^4 x \left( p_a \dot{a} - N {\mathcal{H}}_{grav} \right) 
\end{align}
where the Hamiltonian constraint is
\begin{equation}
{\mathcal{H}}_{grav} = + \frac{1}{12} \frac{p_a^2}{a} + 3 ka.
\end{equation}
The $\phi$- Hamiltonian constraint should be added to the gravitational part. We get by considering spatially homogeneous $\phi$
\begin{equation}
\mathcal{H} = \mathcal{H}_{grav} + \mathcal{H}_{\phi} = + \frac{1}{12} \frac{ p_a^2}{a} + 3 ka + p + a^3 V(\phi).
\end{equation}
To apply the Wheeler-DeWitt quantization scheme, we set
\begin{align}
p_a = -i \frac{\partial}{\partial a}, \quad \quad p = -i \frac{\partial}{\partial \phi} .
\end{align}
This will form the operator $\hat{\mathcal{H}}$ and the Wheeler-DeWitt equation will give us the evolution of the wavefunction of the universe, $\Psi$, by demanding 
\begin{equation}
\hat{\mathcal{H}} \Psi = 0.
\end{equation}
This equation takes the form
\begin{equation}
-\frac{1}{12a} \frac{\partial^2 \Psi}{\partial a^2} + 3ka \Psi - i \frac{\partial \Psi}{\partial \phi} + a^3 V(\phi) \Psi = 0
\end{equation}
and it can be written as a Schrodinger equation after rearranging 
\begin{equation}
i \frac{\partial \Psi}{\partial \phi} = - \frac{1}{12a} \frac{\partial^2 \Psi}{\partial a^2} + 3ka \Psi
\label{wheelereq}
\end{equation}
where $V(\phi)$ was taken to be zero for simplicity. For the operator 
\begin{equation}
\hat{H} = - \frac{1}{12a} \frac{\partial^2}{\partial a^2} + 3ka
\end{equation}
to be self-adjoint, the inner product of any two wave functions is defined by
\begin{equation}
\left( \psi_1, \psi_2 \right) = \int_{0}^{\infty} a \ \psi_1^* \psi_2 da.
\end{equation}
For $k=0$, the operator $\hat{H}$ is symmetric if
\begin{align}
\left( \psi_1, \hat{H} \psi_2 \right) = \left( \hat{H} \psi_1, \psi_2 \right) \nonumber \\
\end{align}
which gives
\begin{align}
\int_{0}^{\infty} \psi_1^* \frac{d^2 \psi_2}{d a^2} \ da = \int_{0}^{\infty} \frac{d^2 \psi_1^*}{d a^2} \psi_2 \ da.
\end{align}
Thus for $\psi$ and its derivative being square-integrable, the simplest boundary conditions imposed to the wave function are
\begin{equation}
\Psi(0,\phi) = 0 \quad \text{or} \quad \frac{\partial \Psi(a,\phi)}{\partial a}\vert_{a=0} = 0
\end{equation}
\paragraph{}
The Wheeler-DeWitt equation (\ref{wheelereq}) can be solved by separation of variables. This will result in stationary solutions of the form
\begin{equation}
\Psi(a,\phi) = e^{- iE \phi} \psi(a),
\end{equation}
and this leads to a differential equation in $\psi(a)$
\begin{equation}
\frac{d^2 \psi}{d a^2} - 36 k a^2 \psi + 12 E a \psi = 0
\end{equation}
where E is a real parameter. \newline
The general solution to the above equation in the case of flat universe (k = 0) is
\begin{equation}
\psi(a) = \sqrt{a} \left[ A J_{1/3} \left( \frac{2}{3} \sqrt{12 E} a^{3/2} \right) + B J_{-1/3} \left( \frac{2}{3} \sqrt{12 E} a^{3/2} \right)  \right] .
\label{bessel}
\end{equation}
Upon applying the first boundary condition and due to the behaviour of the Bessel function, $B$ is set to zero. If instead the second boundary condition is imposed then $A = 0$. \newline
The norm of these solutions is infinite; hence, wave packets must be constructed by superposing the solutions. This can be done similar to the work in \cite{dust} and \cite{dust1} where we have
\begin{equation}
\Psi(a,\phi) = \int^{\infty}_{0} c(E) \ e^{- iE \phi} \psi(R). 
\end{equation}
\paragraph{}
Solutions can also be found to the cases $k = 1$ and $k = -1$. For the first case, the solutions are given in terms of Hermite polynomials while for the latter we get Whittaker functions. This all in agreement to what was found before for dust in \cite{dust}.

\subsection{$V(\phi)$ Being a Constant}

Above we took $V(\phi) = 0$ and our results were all in agreement with what was done before for dust. Now we will look for solutions when the potential takes a constant value, $V(\phi) = \lambda$. The condition of self-adjointness still applies for $\lambda$ being real. \newline
The equation then takes the form
\begin{equation}
\frac{d^2 \psi}{d a^2} + 12 E a \psi - 36 k a^2 \psi - 12 a^4 \lambda \psi  = 0.
\end{equation}
The stationary solutions for $k=0$ are given by Whittaker functions 
\begin{equation}
\Psi(a,\phi) = e^{-i E \phi} a^{-1} \left\lbrace C_1 M_{\frac{E}{ \sqrt{3 \lambda}}, \frac{1}{6}}\left( \frac{4 \sqrt{3\lambda} a^3} {3} \right) + C_2 W_{\frac{E}{ \sqrt{3 \lambda}}, \frac{1}{6}}\left( \frac{4 \sqrt{3\lambda} a^3} {3} \right) \right\rbrace.
\end{equation}
Wave packets are hard to obtain because integrals over Whittaker functions are difficult to deal with. \paragraph{}
These Whittaker functions can be written in terms of hypergeometric functions
\begin{align}
\psi(a) = {}& \frac{C_1}{3 a \ e^{\frac{2}{3} \sqrt{3 \lambda} a^3}} \left( 4^{\frac{2}{3}} 3^{\frac{2}{3}} \lambda^{\frac{1}{3}} a^2 \ \text{hypergeom} \left( \left[ \frac{2}{3} - \frac{E}{\sqrt{3\lambda}} \right] , \left[ \frac{4}{3}\right], \frac{4}{3} \sqrt{3\lambda} a^3 \right) \right) \nonumber \\
& + \frac{2 C_2}{3 e^{\frac{2}{3} \sqrt{3 \lambda} a^3}} \left(\frac{ 4^{\frac{1}{3}} 3^{\frac{1}{3}} \lambda^{\frac{1}{6}} a \pi \ \text{hypergeom} \left( \left[ \frac{1}{3} - \frac{E}{\sqrt{3\lambda}} \right] , \left[ \frac{2}{3}\right], \frac{4}{3} \sqrt{3\lambda} a^3 \right) } {\Gamma\left(\frac{2}{3} \right) \Gamma\left( \frac{2}{3} - \frac{E}{\sqrt{3\lambda}} \right) } \nonumber \parenthnewln
- \frac{ 4^{\frac{2}{3}} 3^{\frac{2}{3}} \lambda^{\frac{1}{3}} a^2 \Gamma\left( \frac{2}{3}\right) \ \text{hypergeom} \left( \left[ \frac{2}{3} - \frac{E}{\sqrt{3\lambda}} \right] , \left[ \frac{4}{3}\right], \frac{4}{3} \sqrt{3\lambda} a^3 \right) } {\Gamma\left( \frac{1}{3} - \frac{E}{\sqrt{3\lambda}} \right) } \right).
\end{align}
Taking the limit $a \rightarrow 0$, we get
\begin{equation}
\psi(a) \vert_{a \rightarrow 0} = \frac{2}{3} \frac{C_2 \ 4^{1/3} 3^{1/3} \lambda^{1/6} \pi}{\Gamma\left(\frac{2}{3} \right) \Gamma\left(\frac{1}{3} - \frac{E}{\sqrt{3\lambda}} \right) }.
\end{equation}
The boundary condition $\Psi(0,\phi) = 0$ is satisfied upon choosing $C_2 = 0$. \newline
To take the limit as $\lambda = 0$, we first write the hypergeometric function as a series. Also, we use the asymptotic series for $\frac{1}{\Gamma(z)}$ which is given by
\begin{equation}
\frac{1}{\Gamma(z)} \sim z + \gamma z + \frac{1}{12} \left( 6 \gamma^2 - \pi^2 \right) z^3 + \frac{1}{12} \left[ 2 \gamma^3 - \gamma \pi^2 + 4 \zeta(3) \right] z^4 + \cdots  
\end{equation}
where $\gamma$ is the Euler-Mascheroni constant and $\zeta(z)$ is the Riemann zeta function. Then as $\lambda \rightarrow 0$, the Bessel solution (equation \ref{bessel}) is recovered as expected. 
\paragraph{}
For $k = 1$ and $k = -1$, the solutions turn out to take the form of Heun functions. They are given respectively by
\begin{align}
\psi(a) {}& = C_1 e^{\frac {\sqrt{3} (2 a^3 \lambda + 9a )} {3 \sqrt{\lambda}}}  HeunT \left( \frac{9 \ 18^{2/3}}{4 \lambda^{4/3}}, - \frac {3\sqrt{3} E}{\sqrt{\lambda}}, \frac{3}{2} \frac{12^{2/3}}{\lambda^{2/3}}, -\frac{1}{3} a \ 2^{2/3} \ (243\lambda)^{1/6}\right) \nonumber \\
& + C_2 e^{ - \frac {\sqrt{3} (2 a^3 \lambda + 9a )} {3 \sqrt{\lambda}}}  HeunT \left( \frac{9 \ 18^{2/3}}{4 \lambda^{4/3}}, \frac {3\sqrt{3} E}{\sqrt{\lambda}}, \frac{3}{2} \frac{12^{2/3}}{\lambda^{2/3}}, \frac{1}{3} a \ 2^{2/3} \ (243\lambda)^{1/6}\right) 
\end{align}
and  
\begin{align}
\psi(a) {}& = C_1 e^{\frac {\sqrt{3} (2 a^3 \lambda + 9a )} {3 \sqrt{\lambda}}}  HeunT \left( \frac{9 \ 18^{2/3}}{4 \lambda^{4/3}}, - \frac {3\sqrt{3} E}{\sqrt{\lambda}}, - \frac{3}{2} \frac{12^{2/3}}{\lambda^{2/3}}, -\frac{1}{3} a \ 2^{2/3} \ (243\lambda)^{1/6}\right) \nonumber \\
& + C_2 e^{ - \frac {\sqrt{3} (2 a^3 \lambda + 9a )} {3 \sqrt{\lambda}}}  HeunT \left( \frac{9 \ 18^{2/3}}{4 \lambda^{4/3}}, \frac {3\sqrt{3} E}{\sqrt{\lambda}}, - \frac{3}{2} \frac{12^{2/3}}{\lambda^{2/3}}, \frac{1}{3} a \ 2^{2/3} \ (243\lambda)^{1/6}\right) 
\end{align}
For $a = 0$, the Heun functions are equal to one. Hence, the first boundary condition, $\Psi(0,\phi) = 0$, is satisfied when the sum of the two constants $C_1 + C_2$ equals zero. Unfortunately, integrals are also difficult to be dealt with. 


\section{Conclusion}
In this paper, we constructed the Hamiltonian of Mimetic Gravity. We showed that the two extra equations of motion are the constraint equation and the conservation of energy-momentum tensor. Then, the Poisson brackets are computed and the theory is shown to be closed. At the end, canonical quantization was performed for a homogeneous and isotropic universe. This was done first for a vanishing potential and then for a constant one. \newline
It was later noticed that a similar work was done in \cite{similar} almost simultaneously.

\begin{acknowledgement}
I would like to thank Professor Ali Chamseddine for useful discussions.
\end{acknowledgement}

\end{document}